\documentclass[10pt]{article}
\usepackage{amsmath}
\usepackage{amssymb}
\begin{document}
\begin{flushright}
KFKI-1984-91\\
HU ISSN 0368 5330\\
{\it ISBN 963 372 282 9}\\
August 1985
\end{flushright}
\vskip 3truecm
\begin{center}
GRAVITATION AND QUANTUMMECHANICAL LOCALIZATION\\ OF MACROOBJECTS
\vskip 1truecm
L. Di\'osi
\vskip .5truecm
Central Research Institute for Physics\\
H-1525 Budapest 114, P.O.B. 49, Hungary
\end{center}
\vskip 1truecm
ABSTRACT

We propose nonlinear Schr\"odinger equation with gravitational self-interacting term.
The separability conditions of Bialynicki-Birula are satisfied in asymptotic sense.
Solitonlike solutions were found.
\vskip 3 truecm

There is an old-established common knowledge that when extending quantummechanical
laws to macroscopic bodies one is confronted, among others, with the following problem.

According to classical physics, in the absence of external forces the
center of mass of a given macroobject either moves uniformly along a straight
line or, in the particular case, rests at a certain point. Unfortunately, the
Schr\"odinger equation of a free particle does not have localized stationary
solutions. Wave-packet solutions which are possibly the best representation
for the free motion of a macroscopic body are not stationary. On the contrary,
the wave-packet corresponding to the c.m. continually widens with the time
thus the position of the c.m. becomes more and more uncertain. At the same
time, experiences show that a macroscopic object always has a well defined
position.

A possible way to walk round this contradiction is to exclude the initial
states which develop measurable spread of c.m. of the given macroobject.
For instance, if the wave-function of the given body of several grams weight
is initially localized in a volume of about $10^{-8}$ cm linear size (or larger)
the quantummechanical spreading of its c.m. will be extremely slow. The initial
position displays no change even for thousands of years, and the wave-packet 
of the c.m. is apparently stationary with very good precision.

People often argue that it is meaningless to suppose that a macroscopic 
body can have more accurate localization than the typical atomic size $10^{-8}$ cm.
Nevertheless, this is only a bad guess. Let us accept that the position of
a single atom is usually smeared in a volume of atomic size. Then we must
conclude that the c.m. position of a group of many atoms will be defined much
more accurately than the position of the single atoms.  

This reasoning shows that the atomic size $10^{-8}$ cm does not give an
absolute limitation for localizing macroobjects. Thus it would be conceivable
to suppose a $1$ milligram macroscopic object with a wave-packet of $10^{-12}$ cm
width. However, this initial width becomes several times larger even in a
few minutes. Hence, quantummechanics woud predict nonstationary behaviour
for the free motion of a macroobject and this anomaly could, in principle,
be detected in certain extreme experiments$^1$.

However, Nature can single out an other possible way for solving the
problem of wave-function localization: we cannot exclude the existence of a 
mechanism which modifies the laws of the quantummechanics for macroscopic
objects. A modified  Schr\"odinger equation will then have localized stationary
solutions describing the state of macroobjects. Such arguments were put
forward in Ref. 2, where nonlinear but local term was added to the Schr\"odinger
equation and solitonlike solutions were found.

In the present work we show that the gravitational interaction possibly
could prevent the unbounded quantummechanical spreading of the c.m. position
of macroobjects, at least in certain quantumstates. If this interaction is
included, it destroys the linearity of quantummechanics$^3$. In the nonrelativistic
case, newtonian gravitation can explicitely be built into the Schr\"odinger
equation. We arrive then at a nonlinear integro-differential equation
possessing solitonlike solutions, the ones we need to describe the well-localized
macroobjects.

A theory, satisfactorily unifying the quantummechanics and the gravitation
in every respect, still has not been found. Here we are going to apply 
the approach of M{\o}ller and Rosenfeld$^{4,5}$:
$$
R_{ab}-\frac{1}{2}g_{ab}R=\frac{8\pi G}{c^4}\cdot\langle\psi\vert{\hat T}_{ab}\vert\psi\rangle
\eqno{(1)}
$$
where $g$ is the metrics, $R_{ab}$ is the Ricci-tensor, $G$ stands for the constant of
Newton and $c$ denotes the velocity of the light. On the RHS of the Einstein 
equation we put the expectation value of the energy-momentum tensor operator
${\hat T}_{ab}$ in the actual quantum state $\psi$.

This equation is certainly not correct if the fluctuation of $T_{ab}$ is too
large in the quantumstate $\varphi$, e.g. when macroscopically different densities
of the energy and momentum are superposed$^6$. But, if the given quantumstate $\psi$ 
can definitely be associated with only one microstate, there can be no a
priori objection against Eq. (1). Actually, this equation is to be applied 
as long as we do not quantize gravity.

Henceforth we shall discuss nonrelativistic systems. Let us consider
the Schr\"odinger equation for a system of $N$ particles having masses $m_1,m_2...,m_N$:
$$
i\hbar\frac{\partial}{\partial t}\psi(X,t)
=\left[-\sum_{r=1}^N\frac{\hbar^2}{2m_r}\frac{\partial^2}{\partial x_r^2}
            +\sum_{r,s=1}^N V_{rs}(x_r-x_s) +\sum_{r=1}^N m_r \Phi(x_r,t)\right]\psi(X,t).
\eqno{(2)}
$$            
Here, $X\equiv(x_1,x_2,...x_N)$ stands for the space coordinates of the particles,
$V_{rs}$ is the interaction potential and $\Phi$ denotes the newtonian gravitational
potential given by the nonrelativistic equivalent of the Einstein equation
(1):
 $$
\Delta\Phi(x,t)=-4\pi G\int d^{3N}\!\!X^\prime \vert\psi(X^\prime,t)\vert^2 \sum_{r=1}^N \delta^{(3)}(x-x_r^\prime).
\eqno{(3)}
$$
If we solve the Poisson equation (3) explicitely, we can eliminate the
potential $\Phi$ from Eq. (2). Thus we are led to the following nonlinear integro-differential
equation:
$$
i\hbar\frac{\partial}{\partial t}\psi(X,t)
=\left[-\sum_{r=1}^N\frac{\hbar^2}{2m_r}\frac{\partial^2}{\partial x_r^2}
            +\sum_{r,s=1}^N V_{rs}(x_r-x_s)\right.~~~~~~~~~~~~~~~~~~~~~~~~~~~~~~~~~~~~
$$
\vskip -.8truecm
$$\eqno{(4)}$$
\vskip -.8truecm
$$ 
~~~~~~~~~~~~~~~~~~~~~~~~~~~  
         \left.-G\sum_{r,s=1}^N\int\frac{m_r m_s}{\vert x_s^\prime-x_r\vert} \vert\psi(X^\prime,t)\vert^2d^{3N}\!\!X^\prime\right]\psi(X,t).
$$      

For one free pointlike object of mass $M$, Eq. (4) reduces to the following
nonlinear Schr\"odinger equation with non-local self-interacting term:
$$
i\hbar\frac{\partial}{\partial t}\psi(x,t)
=-\frac{\hbar^2}{2M}\Delta\psi(x,t) -GM^2\int\frac{\vert\psi(x^\prime,t)\vert^2}{\vert x^\prime-x\vert} d^3\!x^\prime\psi(x,t).
 \eqno{(5)}
 $$

An important feature of Eq. (4) is that it asymptotically satisfies the
separability condition of Bialynicki-Birula$^2$: Let $\psi^{(A)}(x_A,t)$ and  $\psi^{(B)}(x_B,t)$
be solutions to Eq. (5) for single particles $A$ and $B$ respectively. If the
spatial separation of $A$ and $B$ is large enough to neglect both the potential
$V_{AB}$ and the gravitational interaction between $A$ and $B$, then the wave-function
$\psi^{AB}(x_A,x_B,t)=\psi^{(A)}(x_A,t)\cdot\psi^{(B)}(x_B,t)$ is a solution to the two-particle
equation (4) with $N=2$.

Let us remind that in Ref.2 only mathematically local nonlinearities
were discussed. Our nonlinear term is nonlocal.

We note that in Eqs. (4), (5) the wave-functions must be normalized to
the unity. The nonlinear Schr\"odinger equation (5) preserves this normalization:
$$
\frac{d}{dt}\int\vert\psi(x,t)\vert^2 d^3x = 0,
\eqno{(6)}
 $$
and the expectation value of the momentum operator $\hat p$ and that of the energy
operator $\hat E$ are conserved as well:
$$
\frac{d}{dt}\langle\psi\vert{\hat p}\vert\psi\rangle\equiv\frac{d}{dt}\int\psi^{\!\!\!\ast}(x,t)(-i\hbar\nabla)\psi(x,t)d^3x=0,
\eqno{(7)}
$$
$$
\frac{d}{dt}\langle\psi\vert{\hat E}\vert\psi\rangle\equiv
\frac{d}{dt}\int\psi^{\!\!\!\ast}(x,t)\left(-\frac{\hbar^2}{2M}\Delta-\frac{GM^2}{2}\int\frac{\vert\psi(x^\prime,t)\vert^2}{\vert x^\prime-x\vert} d^3\!x^\prime\right)
                             \psi(x,t)d^3x=0.
\eqno{(8)}
$$

Naturally, Eq. (5) is covariant against galilean transformations. It
can be shown that if $\psi(x,t)$ solves the Eq. (5) then the function
$$
\psi(x-r-vt,t)\mathrm{e}{-\frac{i}{\hbar}\frac{Mv^2}{2}t+\frac{i}{\hbar}Mvx\atop}{}
\eqno{(9)}
$$
will also be a solution, where $r$ and $v$ are arbitrary constants.

Certain solutions of unit norm can conveniently describe the quantummechanical
propagation of a given pointlike macroobject of mass $M$. We are
going to show that the solution of minimal energy is a solitonlike fixed
wave-packet with static spatial density. Let us consider the normalized
function $\varphi(x)$ minimizing the energy functional (8):
$$
E=\int\varphi^{\!\!\!\ast}(x,t)\left(-\frac{\hbar^2}{2M}\Delta-\frac{GM^2}{2}\int\frac{\vert\varphi(x^\prime,t)\vert^2}{\vert x^\prime-x\vert} d^3\!x^\prime\right)
                             \varphi(x,t)d^3x=\mathrm{min},
\eqno{(10)}
$$
$$
\int\vert\varphi(x,t)\vert^2 d^3x= 1.
\eqno{(11)}
$$

One can easily verify that the phase of $\varphi$ will not depend on the variable 
$x$ thus we can choose $\varphi(x)$ to be a real function. The resulting minimum
problem is 
$$
\frac{\hbar^2}{2M}\int(\nabla\varphi(x))^2 d^3x-\frac{GM^2}{2}\int\frac{\varphi^2(x^\prime)\varphi^2(x)}{\vert x^\prime-x\vert} d^3\!x^\prime d^3x
-\epsilon\int\varphi^2(x)d^3x =\mathrm{min}.
\eqno{(12)}
$$
where $\epsilon$ is a Lagrange multiplier.

It can be proved that if $\varphi_0(x),\epsilon_0$ satisfy the minimum condition (12)
and also the normalization (11) then the wave-function 
$$
\varphi_0(x,t)(=\varphi_0(x)\mathrm{e}^{-i\epsilon_0 t}
\eqno{(13)}
$$
is a solution to the nonlinear Schr\"odinger equation (5). Indeed, substituting
the ansatz (13) into Eq. (5) one arrives at the nonlinear time-independent
Schr\"odinger equation for $\varphi_0(x)$. This latter equation is the same as the variational
equation corresponding to the minimum problem (12). Thus, function
(13) proves to be the ground-state solution to Eq. (5).

Finally, we have to find the function $\varphi_0(x)$. Let us suppose that $\varphi_0(x)$
is a smooth real function of unit norm,which has a peak with a characteristic 
width $\underline{a}$ at the origin and tends to zero outside this region. We can
qualitatively evaluate the expression (10) of the energy $E$, which is now
depending on the width $\underline{a}$:
$$
E\approx\frac{\hbar^2}{Ma^2}-\frac{GM^2}{a}.
\eqno{(14)}
$$

By minimizing this expression we get the characteristic width $a_0$ of the
ground-state wave-function $\varphi_0(x)$:
$$
a_0\approx\frac{\hbar^2}{GM^3}.
\eqno{(15)}
$$

Hence, this value can be taken as the measure of the quantummechanical
uncertainty in the position of a free pointlike macroscopic object. The expression
(13) is the stationary ground-state wave-function of an object localized
at the origin. Applying galilean transformation (9), one can construct
the stationary wave-function corresponding to arbitrary uniform rectilinear
motion of the object.

In addition to these one-soliton solutions, the nonlinear Schr\"odinger
equation (5), unfortunately, possesses other solutions too. These latter are
associated with quantummechanical states which generally cannot occure in
the world of macroobjects. We do not know precisely how to exclude these paradoxical
solutions from the theory. The most natural idea is to suppose
that certain physical mechanism destroys such states.

Let us demonstrate a typically unphysical two-soliton solution. The
propagation of the given pointlike macroobject is described by two wave-packets
of width about $a_0$. Both of them are normalized to $1/2$. The two wave-packets
are moving around each other as if they were two objects with mass
$M/2$, gravitationally attracting each other.

Formula (15) yields the width of the wave-packet of a free pointlike
macroobject, i.e., the extension of the object is much less than the spread
$a_0$ of its position. Now we estimate the value of $a_0$ for a homogeneous spherical
object of radius $R$ and mass $M$, supposing that $a_0\ll R$. The only change
appears in the interaction term in the functionals (10), (12). The simple
newtonian kernel $-GM^2\vert x^\prime-x\vert^{-1}$ has to be substituted by the effective interaction
potential $V(x^\prime-x)$ of two homogeneous spheres with radius $R$ and mass $M$:
$$
V(x^\prime-x) = -\frac{GM^2}{(4\pi R^3/3)^2}\int\displaylimits_{r<R}d^3r\int\displaylimits_{r^\prime<R}d^3r^\prime\frac{1}{\vert x^\prime+r^\prime-x-r\vert}=
$$
\vskip -.4truecm
$$\eqno{(16)}$$
\vskip -.4truecm
$$
~~~~~~~~=\frac{GM^2}{R}\left(-\frac{6}{5}+\frac{1}{2}\left\vert\frac{x^\prime-x}{R}\right\vert^2 +\mathcal{O}\left(\left\vert\frac{x^\prime-x}{R}\right\vert^3\right)\right).
$$

The characteristic $\underline{a}$-dependence of the energy $E$ is the following:
$$
E\approx\frac{\hbar^2}{Ma^2}-\frac{GM^2}{R}+\frac{GM^2}{R^3}a^2.
\eqno{(17)}
$$

The width $a_0^{(R)}$ of the ground-state wave-packet is given by the minimization
of $E$:
$$
a_0^{(R)}\approx\left(\frac{\hbar^2}{GM^3}\right)^{1/4}\cdot R^{3/4}=a_0^{1/4}\cdot R^{3/4}
\eqno{(18)}
$$
where $a_0$ is the spread of the pointlike object, see formula (15).

We consider formulae (15) and (18) as the main result of this work.
We claim that these expressions define the natural width of the wave-packet
of any macroscopic object. A similar problem of the natural uncertainty in
the orientation of an extended macroobject can be discussed in this frame,
too.

It is interesting to note that, in Ref.1, the same result (15) was
obtained from certain principle of the metrical smearing of the space-time.
For extended objects, the relation $a_0^{(R)}\approx a_0^{1/3}\cdot R^{2/3}$ was derived, which is
not identical with our result (18). However, if a critical size $R_c$ is defined
by the condition $a_0^{(R)}=R_c$, then Ref.1 and formula (18) yield the same
value $R_c\approx10^{-5}$ cm for objects of normal density. In Ref.2 special considerations
are used to estimate the critical size and a value of also about
$10^{-5}$ cm was predicted. Both papers$^{1,2}$ and the present one too, adopt the
idea that a breakdown of the superposition principle is foreseen in the macroworld
and $R_c$ defines the line of demarcation between micro- and macroobjects.

\parskip 0truecm
\vskip 1.5 truecm
\noindent
REFERENCES
\vskip .2truecm
\noindent $^{1.}$ F. K\'arolyh\'azy, A. Frenkel, B. Luk\'acs, in: Physics as Natural Philosophy,
edited by A. Shimony and H. Feshbach (The MIT Press, Cambridge, Massachusetts, USA, 1982)
\vskip .2 truecm
\noindent $^{2.}$ I. Bialynicki-Birula, J. Mycielski: Ann. Phys. \underline{100}, 62 (1976)
\vskip .2 truecm
\noindent $^{3.}$ T.W.B. Kibble: Commun.Math.Phys. \underline{64}, 73 (1978)
\vskip .2 truecm
\noindent $^{4.}$ C. M{\o}ller, in: Les Theories Relativistes de la Gravitation, edited by
A. Lichnerowich and M.A. Tonnelat (CNRS, Paris, 1962)
\vskip .2 truecm
\noindent $^{5.}$ L. Rosenfeld, Nucl.Phys. \underline{40}, 353 (1963)
\vskip .2 truecm
\noindent $^{6.}$ D.N. Page, C.D. Geilker, Phys.Rev.Lett. \underline{47}, 979 (1981)

\hskip -5pt B. Hawkins, Phys.Rev.Lett. \underline{48}, 520 (1982)

\hskip -5pt L.E. Ballentine, Phys.Rev.Lett. \underline{48}, 522 (1982)
\end{document}